\documentclass[conference]{IEEEtran}
\IEEEoverridecommandlockouts

\usepackage{cite}
\usepackage{multirow}
\usepackage{amsmath,amssymb,amsfonts,subcaption,hyperref}
\usepackage[linesnumbered,ruled,vlined]{algorithm2e}
\usepackage{booktabs} 
\usepackage{graphicx}
\usepackage{textcomp}
\usepackage{xcolor}
\def\BibTeX{{\rm B\kern-.05em{\sc i\kern-.025em b}\kern-.08em
    T\kern-.1667em\lower.7ex\hbox{E}\kern-.125emX}}

\begin{document}

\title{Integrating Deep Unfolding with Direct Diffusion Bridges for Computed Tomography Reconstruction}

\author{\IEEEauthorblockN{Herman Verinaz-Jadan\IEEEauthorrefmark{1} and Su Yan\IEEEauthorrefmark{2}}
\IEEEauthorblockA{\IEEEauthorrefmark{1}Faculty of Electrical and Computer Engineering (FIEC)\\ Escuela Superior Politecnica del Litoral (ESPOL), Ecuador\\
Email: hverinaz@espol.edu.ec}
\IEEEauthorblockA{\IEEEauthorrefmark{2}Department of Bioengineering,	Imperial College London, UK\\
Email: s.yan18@imperial.ac.uk}
}

\maketitle

\begin{abstract}
Computed Tomography (CT) is widely used in healthcare for detailed imaging. However, Low-dose CT, despite reducing radiation exposure, often results in images with compromised quality due to increased noise. Traditional methods, including preprocessing, post-processing, and model-based approaches that leverage physical principles, are employed to improve the quality of image reconstructions from noisy projections or sinograms. Recently, deep learning has significantly advanced the field, with diffusion models outperforming both traditional methods and other deep learning approaches. These models effectively merge deep learning with physics, serving as robust priors for the inverse problem in CT. However, they typically require prolonged computation times during sampling. This paper introduces the first approach to merge deep unfolding with Direct Diffusion Bridges (DDBs) for CT, integrating the physics into the network architecture and facilitating the transition from degraded to clean images by bypassing excessively noisy intermediate stages commonly encountered in diffusion models. Moreover, this approach includes a tailored training procedure that eliminates errors typically accumulated during sampling. The proposed approach requires fewer sampling steps and demonstrates improved fidelity metrics, outperforming many existing state-of-the-art techniques.
\end{abstract}


\section{Introduction}
\label{sec:intro}

Computed Tomography (CT) is a widely used technique in medicine. Low-dose CT (LDCT) specifically reduces patient exposure to ionising radiation, thereby minimising health risks. However, the reduced radiation dose often results in images with increased noise and artifacts, potentially compromising diagnostic accuracy \cite{xia2023physics,gao2023corediff}.

To mitigate these effects, various techniques have been developed to reconstruct clean images from raw CT data. Traditional methods include sinogram preprocessing, aimed at improving data quality before reconstruction, and post-processing techniques that refine image quality, typically starting with an initial reconstruction obtained via filtered back projection (FBP) \cite{balda2012ray,willemink2019evolution,xia2023physics,9468943}. Furthermore, model-based methods incorporate physical principles and employ regularization strategies such as Total Variation (TV), wavelet-based methods, and sparsity-driven approaches to enhance the accuracy of the reconstruction \cite{chen2013improving}.

Deep learning methods have significantly enhanced CT image reconstruction. Various architectures, such as convolutional neural networks (CNNs) \cite{xia2021ct}, generative adversarial networks (GANs) \cite{jiang2023low}, and transformers \cite{wang2023ctformer}, have pushed the boundaries of CT imaging. While powerful, many of these methods can result in over-smoothed images, lack generalisation, require complex architectures, or are prone to instability during training, presenting significant challenges for clinical applications \cite{gao2023corediff}. To address these issues, models that integrate physical principles into the architecture have been proposed, improving generalisation capabilities and reconstruction fidelity\cite{xiang2021fista,adler2018learned,chen2018learn}.

Recently, diffusion models \cite{song2021solving,ho2020denoising} have emerged as robust tools for synthesising high-quality images. These models serve as effective priors for solving inverse imaging problems. Although they typically require significant computational resources, methods such as Cold Diffusion and Direct Diffusion Bridges (DDBs) have been developed to facilitate faster transitions from degraded to high-quality images \cite{bansal2024cold,delbracio2023inversion,10.5555/3618408.3619323}.

This paper introduces a novel framework that enhances CT reconstruction by integrating system physics with diffusion models. Leveraging a DDB and a deep unfolding network that is naturally compatible with diffusion processes, this framework includes physics in the architecture rather than in the sampling stage, unlike previous approaches \cite{nichol2021improved,song2020denoising,bansal2024cold,karras2022elucidating,gao2023corediff,ho2020denoising}. The proposed method reconstructs images from raw sinograms with high fidelity, aligns closely with the physical model, and reduces the need for extensive sampling.
\section{Direct Diffusion Bridge}
A DDB is a novel diffusion approach to image reconstruction that simplifies direct transition from a degraded to a clean image, bypassing excessively noisy intermediate stages. Unlike traditional diffusion processes, which gradually denoise starting from pure noise, DDBs specifically target the reconstruction process~\cite{NEURIPS2023_165b0e60,ho2020denoising}.

Given the initial state, $\mathbf{x}_0$, representing the original image sampled from the data distribution $p_0$, and a degraded version, $\mathbf{x}_1$, sampled from $p_{1|\mathbf{x}_0}$, the intermediate state $\mathbf{x}_t$ at any time $t$ can be described with a Gaussian mixture~\cite{10.5555/3618408.3619323,NEURIPS2023_165b0e60}:
\begin{equation}
p(\mathbf{x}_t|\mathbf{x}_0, \mathbf{x}_1) = \mathcal{N}\left(\mathbf{x}_t; (1-\alpha_t)\mathbf{x}_0 + \alpha_t\mathbf{x}_1, \sigma_t \mathbf{I}\right). 
\label{eq:DDBIntSt} 
\end{equation}

This formulation is implemented using the reparameterization trick, enabling sampling by:
\begin{equation} 
\mathbf{x}_t = (1 - \alpha_t)\mathbf{x}_0 + \alpha_t\mathbf{x}_1 + \sigma_t\mathbf{z}, \quad \mathbf{z} \sim \mathcal{N}(0, \mathbf{I}). 
\label{eq:DDBIntStRT} 
\end{equation}

Experimentally, a time-dependent neural network is leveraged during sampling to map noisy intermediate states to the clean image $\mathbf{x}_0$ at each step, as detailed in~\cite{10.5555/3618408.3619323,NEURIPS2023_165b0e60,delbracio2023inversion}.

\subsection{Parameter selection}
In our discussion, we focus on the Image-to-Image Schrodinger Bridge (I²SB) framework based on Schrödinger bridges~\cite{Leonard2014,schrodinger1932theorie}, but other strategies can be employed~\cite{NEURIPS2023_165b0e60,delbracio2023inversion,10.5555/3618408.3619323}. Specifically, I²SB proposes the following forward stochastic differential equation:
\begin{equation}\label{eq:DBEq}
dx_t = \left[ f_t + \beta_t \nabla \log \Psi(x_t, t) \right] dt + \sqrt{\beta_t} dw_t,
\end{equation}
where $ x_0 \sim p_0 $ and $ x_1 \sim p_1 $ and $\Psi \in C^{2,1}(\mathbb{R}^d, [0, 1])$. Under some assumptions~\cite{10.5555/3618408.3619323}, the posterior $p(\mathbf{x}_t|\mathbf{x}_0, \mathbf{x}_1)$ can be computed as in Equation~\eqref{eq:DDBIntSt}. Specifically, it can be shown that the mixing coefficients and variance are defined as follows:
\begin{equation}\label{eq:coffVarI2SB}
\alpha_t = \frac{\gamma_t^2}{\gamma_t^2 + \tilde{\gamma}_t^2}, \quad \sigma_t^2 = \frac{\gamma_t^2 \tilde{\gamma}_t^2}{\gamma_t^2 + \tilde{\gamma}_t^2},
\end{equation}
where \(\gamma_t\) and \(\tilde{\gamma}_t\) are computed from the integrals:
\begin{equation}\label{eq:gammasI2SB}
\gamma_t^2 = \int_0^t \beta_r \, dr, \quad \tilde{\gamma}_t^2 = \int_t^1 \beta_r \, dr,
\end{equation}
and \(\beta_t\) is a predefined noise schedule. A specific schedule is proposed in~\cite{10.5555/3618408.3619323}:
\begin{equation}\label{eq:betasVarI2SB}
\beta_t = 
\begin{cases} 
\beta_{\min} + 2(\beta_{\max}-\beta_{\min}) t, & t \in [0, 0.5] \\
\beta_{\max} - 2(\beta_{\max}-\beta_{min})(t-0.5), & t \in [0.5, 1.0]
\end{cases}
\end{equation}
with \(\beta_{\min}\) and \(\beta_{\max}\) being adjustable parameters.

\section{Problem Formulation}

CT aims to reconstruct detailed images from projections obtained at multiple angles. Specifically, this work focuses on reconstructing a two-dimensional (2D) image, represented as $\mathbf{x} \in \mathbb{R}^{N \times N}$, where each pixel corresponds to the attenuation coefficient of a specific area in the human body~\cite{li2023learning}. The projection data or sinogram, denoted by $\mathbf{y} \in \mathbb{R}^{N_v \times N_d}$, is obtained through the Radon transform~\cite{radon20051}. Here, $N_v$ and $N_d$ represent the number of projection views and the number of detectors, respectively. After vectorization, the image formation process can be described by a linear model as follows:

\begin{equation}\label{eq:frwdMod}
\mathbf{y} = \mathbf{H}\mathbf{x} + \mathbf{n},
\end{equation}
where $\mathbf{H}$ is the measurement matrix that encapsulates the scanning geometry, and $\mathbf{n}$ includes the noise terms accounting for quantum and electronic noise. In this work, the sinogram $\mathbf{y}$ is obtained via LDCT, characterised by higher noise levels compared to NDCT. In practice, noise can be artificially introduced to simulate LDCT from NDCT images as described in \cite{xia2023physics}, providing pairs of images that can be used for training in deep learning methods.

\section{Network Architecture}
Model unfolding, introduced in~\cite{gregor2010learning}, enhances neural network designs by incorporating the principles of model-based algorithms. A prime example, the Learned Iterative Shrinkage-Thresholding Algorithm (LISTA), transforms the iterative steps of the Iterative Shrinkage-Thresholding Algorithm (ISTA)~\cite{daubechies2004iterative} into discrete network layers, with each layer simulating one iteration. This method, which alternates between gradient descent and a proximal operator, has influenced network architecture design across various domains \cite{xiang2021fista,xia2023physics,9428249,zhang2018ista,zhang2020deep,verinaz2023physics}. Our approach builds on these principles by integrating both gradient step and proximal operator modules, facilitating the modelling of diffusion-like processes.

\begin{figure}[]

(a)
\begin{minipage}{1.0\linewidth}
  \centering
  \centerline{\hspace*{2.7 cm}\includegraphics[trim=100 315 70 205, clip, width=11.0 cm]{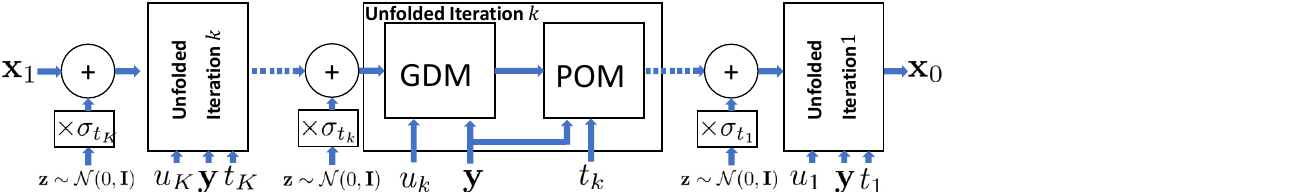}}
\end{minipage}
(b)
\begin{minipage}{1.0\linewidth}
  \centering
  \centerline{\hspace*{0.1 cm}\includegraphics[trim=130 280 300 150, clip, height=4 cm]{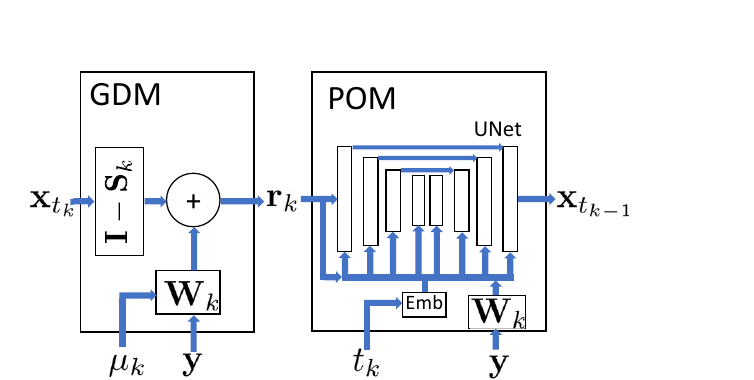}}
\end{minipage}

\caption{Overview of the deep learning model. Part (a) displays the deep unfolding network, while part (b) presents one unfolded iteration formed by the gradient descent module (GDM) and proximal operator module (POM).}

\label{fig:unfArch}
\end{figure}

\subsection{Gradient Descent Module (GDM)}

In this module, the CT model $\mathbf{H}$, the observed degraded image $\mathbf{y}$, and a time-dependent parameter $\mu_k$ interact to refine the image estimate $\mathbf{x}_{t_{k+1}}$ through a gradient step:
\begin{equation}\label{eq:GDM}
    \mathbf{r}_k = \mathbf{x}_{t_{k+1}} - \mu_k \mathbf{W}^T (\mathbf{H}\mathbf{x}_{t_{k+1}} - \mathbf{y}),
\end{equation}

where $\mathbf{r}_k$ is a intermediate reconstruction and $t_{k+1}$ represents the time step.  The index in $\mathbf{r}_k$ is lower than in $t_{k+1}$, indicating a reverse order to align with the diffusion formulation. In this work, by definition $t_{K}=1$ and $t_{0}=0$. Additionally, $\mathbf{W}^T$ serves a role similar to $\mathbf{H}^T$ in standard gradient descent. In model-based networks, $\mathbf{W}$ allows independent optimisation, as in the original LISTA approach~\cite{gregor2010learning}, or it can be precomputed~\cite{liu2018alista}. In this work, we opt to precompute $\mathbf{W}$ to reduce computational requirements, and we only learn the parameter $\mu_k$, as also proposed in~\cite{xiang2021fista}.

\subsection{Proximal Operator Module (POM)}
Following the gradient descent, the POM employs nonlinear transformations to further refine the image, like a proximal operator. This module is modelled as a time-dependent U-Net:
\begin{equation}\label{eq:POM}
    \mathbf{x}_{t_{k}} = U(\mathbf{r}_k, t_{k},\mathbf{y}),
\end{equation}
where $r_k$ is the output of the GDM, and $U$ denotes a U-Net. Specifically, we leverage the U-Net architecture proposed for CT in \cite{gao2023corediff}. This U-Net allows for conditioning on additional inputs beyond time. We condition the network on the input $\mathbf{r}_k$ and the backprojection $\mathbf{W}^T\mathbf{y}$. See Figure~\ref{fig:unfArch} (a).

\subsection{Integration}

The network architecture integrates the GDM and the POM into a multi-layered structure. The input to the network is $\mathbf{y}$ and $\mathbf{x}_1$,  the latter of which can be set to a coarse reconstruction such as an FBP estimate or the backprojection. This configuration, denoted as $G$, is depicted in Figure~\ref{fig:unfArch} (a).

\begin{algorithm}[t]
\caption{Training}
\label{alg:trainingAlg}
\DontPrintSemicolon
\SetAlgoLined 
\SetKwRepeat{Do}{do}{while} 
\KwIn{$p_{0}, p_{1}, p_{\mathbf{y}}, K, \{\alpha_{t_k}\}_{k=1}^K, \{\sigma_{t_k}\}_{k=1}^K$}
\KwOut{$G^*$}

\Do{not converged}{
    $\mathbf{x}_0 \sim p_{0}$\;
    $\mathbf{x}_1 \sim p_{1|\mathbf{x}_0}$\;
    $\mathbf{y} \sim p_{1|\mathbf{x}_0}$\;

    $k \sim \text{Uniform}\{2, \ldots, K\}$\;
    \tcp{with no grad}
    \Indp 
    \For{$i = K$ \KwTo $2$}{
       $\mathbf{\tilde{x}}_{t_{i}} \gets G_{\theta_{i}}(\mathbf{x}_{t_{i}},\mathbf{y},t_i))$\;
       $z \sim \mathcal{N}(0, 1)$\;
       $\mathbf{x}_{t_{i-1}} \gets \mathbf{\tilde{x}}_{t_{i}} + \sigma_{t_{i}} z$\;
    }
    \Indm 
    $\mathbf{x'}_{t_{k-1}} \gets (1-\alpha_{t_{k-1}}) \mathbf{x_0} +\alpha_{t_{k-1}} \mathbf{x}_1$\;
    Take gradient descent step as in Equation~\eqref{eq:GradEq}
}
 \Return{$G^*$}
\end{algorithm}

\section{Network Training and Sampling}
\subsection{Training}
We introduce a novel training procedure for the unfolded network, leveraging DDBs. Each iteration of the network aims to incrementally refine the estimate by minimising the loss function:
\begin{equation}\label{eq:loss1norm}
 \left\| \mathbf{x}'_{t_{k-1}} - G_{\theta_{k}}(\mathbf{x}_{t_k},\mathbf{y},t_k) \right\|^2,
\end{equation}
where $G_{\theta_{k}}$ represents the combined operation of the GDM and POM at the $k$-th layer and $\mathbf{x}'_{t_{k-1}}$ is a convex combination of the degraded image $\mathbf{x}_1$ and the corresponding clean image $\mathbf{x}_0$, as in line 11 in Algorithm~\ref{alg:trainingAlg}. Unlike previous DDBs~\cite{delbracio2023inversion, ho2020denoising, 10.5555/3618408.3619323}, we do not estimate $\mathbf{x}_0$ at every step. Instead, our network is designed to make small, incremental improvements at each unfolded iteration.

Furthermore, we found it beneficial to explicitly include the term for $t_1$ in the loss. As also noted in previous works \cite{zheng2024non,nichol2021improved}, training for low degradation levels can have significant impact on performance. Therefore, during training, the gradient step is taken as follows:
\begin{equation}\label{eq:GradEq}
\nabla_\theta \left(\left\| \mathbf{x}'_{t_{k-1}} - G_{\theta_{k}}(\mathbf{x}_{t_k},\mathbf{y},t_k) \right\|^2 + \left\| \mathbf{x}_{0} - G_{\theta_{1}}(\mathbf{x}_{t_1},\mathbf{y},t_1) \right\|^2\right),
\end{equation}

\begin{algorithm}[!t]
\caption{Sampling}
\label{alg:samplingAlg}
\DontPrintSemicolon
\KwIn{$\mathbf{y}, \mathbf{x}_1, G^*, K, \{\sigma_{t_k}\}_{k=1}^K$}
\KwOut{$\mathbf{x}_0$}

\For{$k = K$ \KwTo $1$}{
    $z \sim \mathcal{N}(0, 1)$\;
    $\mathbf{x}_{t_{k-1}} \gets G_{\theta_{k}}^*(\mathbf{x}_{t_k},\mathbf{y},t_k)+\sigma_{t_{k}} z$\;
 }
 \Return{$\mathbf{x}_0$}
\end{algorithm}
\setlength{\textfloatsep}{8pt}

where $ k $ is randomly selected from 2 to $ K $. For any $k$, $\mathbf{x}_{t_k}$ is obtained by iteratively evaluating the network layers from $K$ down to $k+1$, as detailed in Algorithm~\ref{alg:trainingAlg}.

In previous works \cite{gao2023corediff, bansal2024cold}, $\mathbf{x}_{t_k}$ is estimated as a convex combination of $\mathbf{x}_0$ and $\mathbf{x}_1$. However, discrepancies during sampling lead to cumulative errors that impair performance. To mitigate this issue, coupling stages and adapted sampling techniques have been suggested in \cite{gao2023corediff} and \cite{bansal2024cold}, respectively. In contrast, the approach outlined in Algorithm~\ref{alg:trainingAlg} successfully avoids such errors. Notably, the loop in our method does not require gradient computation, and the number of iterations, $K$, is smaller than typically observed in DDPM models \cite{ho2020denoising}, which helps reduce computational demands.

\subsection{Sampling}

For sampling, as opposed to Equation~\eqref{eq:DDBIntStRT}, the convex combination of $\mathbf{x}_0$ and $\mathbf{x}_1$ is incorporated within the optimised $G^*$. Thus, it is only necessary to iteratively evaluate each stage of $G^*$ while adding noise, as detailed in Algorithm \ref{alg:samplingAlg}. Since we use a similar routine to create the initial estimate in the training Algorithm \ref{alg:trainingAlg}, the accumulation of errors that could impair network performance is avoided.

\section{ Experiments and results}
\label{sec:format}
\begin{figure}[t]
\begin{minipage}{1.0\linewidth}
  \centering
  \centerline{\hspace*{0 cm}\includegraphics[trim=0 0 0 0, clip, width=8.8 cm]{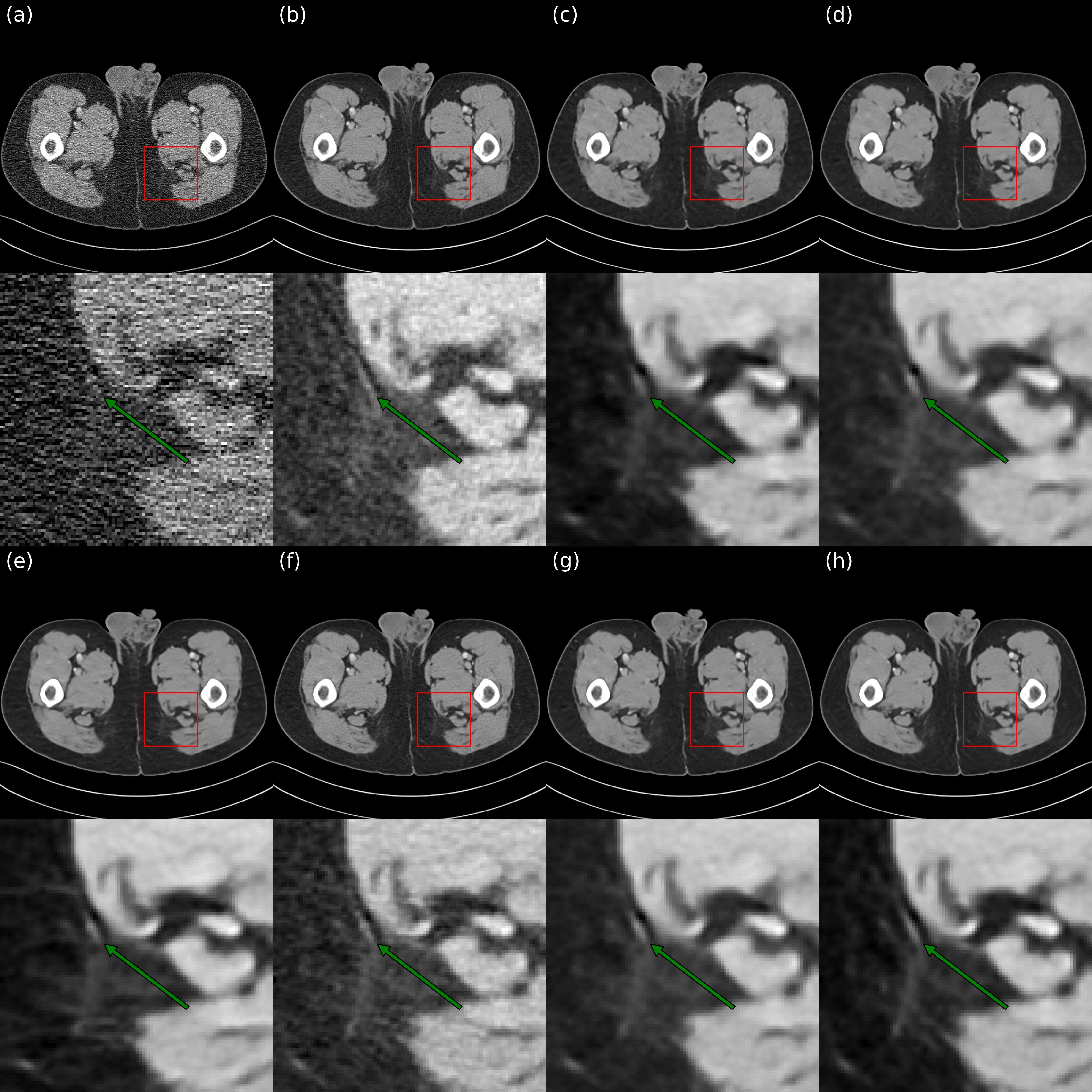}}
\end{minipage}

\caption{Reconstruction results. (a)
LDCT (FBP)~\cite{xia2023physics} (b) NDCT (Ground Truth) (c) LPD~\cite{adler2018learned} (d) FistaNet~\cite{xiang2021fista}  (e) AdaptiveNet~\cite{ge2020adaptive} (f) CoreDiff~\cite{gao2023corediff} (g) LEARN~\cite{chen2018learn} (h) Ours. The CT display window is [-160,240] HU. 
}
\label{fig:resultCT}
\end{figure}

We use the publicly available ``2016 NIH-AAPM-Mayo Clinic Low Dose CT Grand Challenge" dataset \cite{mccollough2017low}, comprising 2378 CT slices (3 mm thickness, 512 × 512 resolution) from 10 patients, as outlined in \cite{xia2023physics}. Then, $N=N_v=512$ and $N_d=512$. We set 600 training images from 7 patients, 100 validation images from one patient, and 200 test images from two patients. LDCT noise levels were set at 20\% of NDCT photon counts with an electronic noise variance of 8.2. For methods requiring forward models, projections were simulated using a distance-driven approach, as in \cite{xia2023physics}.

Next, we detail the specific parameters and configurations used in our experiments. For $\beta_t$, we follow the schedule in Equation~\eqref{eq:coffVarI2SB}, with $\beta_{\text{min}} = 1 \times 10^{-8}$ and $\beta_{\text{max}} = 3.005 \times 10^{-6}$. The noise level, defined by $\sigma_t$ in Equation~\eqref{eq:coffVarI2SB}, applies to both training and sampling, unless stated otherwise. The architecture comprises $K=6$ layers and $\mathbf{x}_1$ is the FBP reconstruction. We use the UNet model also proposed in \cite{gao2023corediff}, but conditioning it on $\mathbf{r}_k$ and $\mathbf{W}^T\mathbf{y}$, as depicted in Figure \ref{fig:unfArch}~(b). For all methods, we trained for $200$ epochs with a batch size of $4$. The models were optimised using the AdamW optimiser with a learning rate of $1 \times 10^{-4}$, maintaining default settings for other parameters. To ensure a fair comparison, CoreDiff~\cite{gao2023corediff} is implemented with the context variable set to zero, and all methods are configured as specified in \cite{xia2023physics}.

\begin{table}[t]
\centering
\caption{Performance Comparison (Mean $\pm$ SD)}
\label{tab:results1}

\begin{tabular}{@{}lccc@{}}
\toprule
Method & PSNR (dB) $\uparrow$ & SSIM $\uparrow$ & LPIPS $\downarrow$ \\
\midrule
FBP~\cite{xia2023physics}                & 33.52 $\pm$ 1.86 & 0.7232 $\pm$ 0.0778 & 0.2708 $\pm$ 0.0595 \\
AdaptiveNet~\cite{ge2020adaptive}        & 43.64 $\pm$ 1.36 & 0.9731 $\pm$ 0.0102 & 0.0632 $\pm$ 0.0341 \\
CoreDiff~\cite{gao2023corediff}           & 43.42 $\pm$ 1.35 & 0.9717 $\pm$ 0.0102 & \textbf{0.0283 $\pm$ 0.0170} \\

LEARN~\cite{chen2018learn}               & 43.60 $\pm$ 1.36 & 0.9732 $\pm$ 0.0102 & 0.0630 $\pm$ 0.0312 \\
LPD~\cite{adler2018learned}                & 43.17 $\pm$ 1.33 & 0.9710 $\pm$ 0.0106 & 0.0582 $\pm$ 0.0289 \\
FistaNet~\cite{xiang2021fista}           & 43.57 $\pm$ 1.35 & 0.9729 $\pm$ 0.0102 & 0.0621 $\pm$ 0.0321 \\
Ours               & \textbf{43.74 $\pm$ 1.39} & \textbf{0.9739 $\pm$ 0.0101} & 0.0614 $\pm$ 0.0359 \\
\bottomrule
\vspace{-0.3cm}
\end{tabular}
\end{table}

Table~\ref{tab:results1} contrasts various reconstruction methods, starting with FBP as a baseline, followed by two pure deep learning methods, and concluding with deep unfolding methods. Our approach achieves the best performance in terms of fidelity, as measured by PSNR and SSIM metrics. It also performs competitively in perceptual quality, assessed using the Learned Perceptual Image Patch Similarity (LPIPS) \cite{zhang2018unreasonable}, although CoreDiff records the highest score. All metrics are computed within the window range [-1000,1000] HU.

Visual results for all methods are presented in Fig \ref{fig:resultCT}. Our method shows improved clarity in certain structures, as indicated by a green arrow, where it appears sharper compared to others. Observe that in this experiment CoreDiff does not reduce noise relative to the NDCT image. This results in reduced fidelity (PSNR and SSIM) but achieves better LPIPS values, as detailed in Table \ref{tab:results1}.
\begin{table}[t]
\centering
\caption{Ablation: Sampling Noise}
\label{tab:sigma_ablation_study}

\begin{tabular}{@{}lcccc@{}}
\toprule
Sigma & PSNR (dB) $\uparrow$ & SSIM $\uparrow$ & LPIPS $\downarrow$ \\ \midrule
$\sigma_t$ & \textbf{43.74} $\pm$ 1.39 & \textbf{0.9739} $\pm$ 0.0101 & 0.0614 $\pm$ 0.0359 \\
$3\sigma_t$ & 43.72 $\pm$ 1.39 & 0.9738 $\pm$ 0.0100 & 0.0612 $\pm$ 0.0359 \\
$6\sigma_t$ & 43.67 $\pm$ 1.37 & 0.9735 $\pm$ 0.0100 & 0.0606 $\pm$ 0.0359 \\
$9\sigma_t$ & 43.58 $\pm$ 1.34 & 0.9730 $\pm$ 0.0099 & 0.0596 $\pm$ 0.0359 \\
$12\sigma_t$ & 43.46 $\pm$ 1.30 & 0.9724 $\pm$ 0.0098 & 0.0583 $\pm$ 0.0358 \\
$15\sigma_t$ & 43.30 $\pm$ 1.24 & 0.9715 $\pm$ 0.0096 & \textbf{0.0566} $\pm$ 0.0357 \\
\bottomrule

\end{tabular}
\end{table}
\begin{table}[t]
\centering
\caption{Ablation: Loss Functions and GDM}
\label{tab:loss_functions}
\begin{tabular}{@{}c@{\hskip 5pt}c@{\hskip 5pt}c@{\hskip 10pt}c@{\hskip 10pt}c@{\hskip 10pt}c@{}}
\toprule
$L_1$ & $L_2$ & GDM & PSNR (dB) $\uparrow$ & SSIM $\uparrow$ & LPIPS $\downarrow$ \\ 
\midrule
\checkmark &  &  & 43.38 $\pm$ 1.35 & 0.9722 $\pm$ 0.0106 & 0.0705 $\pm$ 0.0411 \\
 & \checkmark &  & 43.61 $\pm$ 1.40 & 0.9732 $\pm$ 0.0102 & 0.0627 $\pm$ 0.0367 \\
\checkmark & \checkmark &  & 43.67 $\pm$ 1.38 & 0.9736 $\pm$ 0.0101 & 0.0616 $\pm$ 0.0369 \\
\checkmark &  & \checkmark & 43.65 $\pm$ 1.38 & 0.9733 $\pm$ 0.0103 & 0.0675 $\pm$ 0.0389 \\
 & \checkmark & \checkmark & 43.64 $\pm$ 1.41 & 0.9732 $\pm$ 0.0103 & 0.0616 $\pm$ 0.0369 \\
\checkmark & \checkmark & \checkmark & \textbf{43.74 $\pm$ 1.39} & \textbf{0.9739 $\pm$ 0.0101} & \textbf{0.0614 $\pm$ 0.0359} \\
\bottomrule
\end{tabular}
\end{table}

\begin{table}[h!]
\centering
\caption{Ablation: Number of unfolded iterations ($K$)}
\label{tab:ablation_study}

\begin{tabular}{@{}lccc@{}}
\toprule
K & PSNR (dB) $\uparrow$ & SSIM $\uparrow$ & LPIPS $\downarrow$ \\ \midrule
5 & 43.70 $\pm$ 1.39 & 0.9737 $\pm$ 0.0101 & 0.0613 $\pm$ 0.0361 \\
6 & \textbf{43.74 $\pm$ 1.39} & \textbf{0.9739 $\pm$ 0.0101} & 0.0614 $\pm$ 0.0359 \\
7 & 43.74 $\pm$ 1.39 & 0.9738 $\pm$ 0.0100 & \textbf{0.0612 $\pm$ 0.0357} \\
8 & 43.69 $\pm$ 1.39 & 0.9736 $\pm$ 0.0101 & 0.0616 $\pm$ 0.0361 \\
9 & 43.71 $\pm$ 1.39 & 0.9737 $\pm$ 0.0101 & 0.0617 $\pm$ 0.0363 \\
\bottomrule
\vspace{-0.4 cm}
\end{tabular}
\end{table}

An ablation study on the level of sampling noise is presented in Table~\ref{tab:sigma_ablation_study}. It was found that slight increases in $\sigma_t$ during sampling enhance perceptual quality for this dataset, as evidenced by improved LPIPS scores shown in Table~\ref{tab:sigma_ablation_study}. This allow to achieve the second-best performance among the methods listed in Table~\ref{tab:ablation_study}. Yet, further increases in $\sigma_t$ lead to larger reductions in PSNR and SSIM, ultimately resulting in a degradation of perceptual quality.

The integration of the proposed loss functions and physics principles enhances reconstruction performance. Removing the GDM or employing a single term in Equation~\eqref{eq:GradEq} reduces the effectiveness of our method, as detailed in the ablation study in Table~\ref{tab:loss_functions}. The combination of both loss terms with the GDM achieves the best results in terms of PSNR, SSIM, and LPIPS metrics. Here, $L_1$ represents the first term and $L_2$ the second term in the gradient Equation~\eqref{eq:GradEq}.

Finally, we examined the effect of the number of unfolded iterations on performance, as shown in Table~\ref{tab:ablation_study}. Optimal performance in terms of PSNR and SSIM was found at $ K = 6 $, while $ K = 7 $ yielded the best results for LPIPS, slightly better than $K = 6 $. We chose $K = 6 $ to balance performance with computational efficiency.

\section{Conclusions}
\label{sec:Conclusions}
This work introduces a pioneering framework for CT reconstruction that integrates system physics with a Direct Diffusion Bridge via deep unfolding. The proposed method enhances fidelity metrics such as PSNR and SSIM and delivers competitive perceptual quality, measured by LPIPS, as demonstrated on the Mayo dataset. Furthermore, it addresses the computational demands of sampling in diffusion models. Future work will explore the application of our approach to other inverse problems in imaging.

\bibliographystyle{IEEEbib.bst}
{\bibliography{refs.bib}}

\end{document}